\documentstyle[twoside,fleqn,espcrc2,epsf]{article}

\newcommand{\beq}{\begin{equation}}
\newcommand{\eeq}{\end{equation}}

\title{Dynamics of the Conformal Mode and Simplicial Gravity}

\author{S. Catterall\address{Physics Department, 
        Syracuse University, \\ 
        Syracuse, NY 13244}%
        , E. Mottola and T. Bhattacharya
        \address{Theoretical Division T-8, Mail Stop B285, \\
        Los Alamos National Laboratory, Los Alamos, NM 87545}}
       
\begin{document}

\begin{abstract}
We review the derivation of the Liouville action in 2DQG via the trace anomaly
and emphasize how a similar approach can be used to derive an effective
action describing the long wavelength dynamics of the conformal
factor in 4D. In 2D we describe how to make an explicit connection
between dynamical triangulations and this continuum theory, and present
results which confirm the equivalance of the two approaches. By
reconstructing a lattice conformal mode from 
DT simulations it should be possible to test this
equivalence in 4D also.
\end{abstract}

\maketitle

\section{Quick Review of 2DQG}
The usual FPI approach to 2DQG starts from the partition function
\beq
Z=\int \frac{Dg}{\rm Vol(Diffs)} e^{-S(g)}
\eeq
The measure is constructed to sum only over physically inequivalent
metrics. Choosing the conformal gauge $g=\overline{g}e^\phi$ and implementing
the usual FP procedure leads to the following form for $Z$
\beq
Z\sim \int D\phi e^{-S_L\left(\phi\right)}\times \left(\rm ghosts\right)
\eeq
where the celebrated Liouville action $S_L\left(\phi\right)$ is given by
\beq
S_L\left(\phi\right)=\frac{\left(25-c_M\right)}{96\pi}\int \sqrt{\overline{g}}\left(\overline{g}^{\alpha\beta}\partial_\alpha\phi\partial_\beta\phi+2\overline{R}\phi\right)
\eeq
The parameter $c_M$ is the central charge of massless matter fields.
An entirely equivalent way to derive this action proceeds via the trace
anomaly of the energy momentum tensor of
massless fields in a curved background \cite{emil}. The form of this is

\begin{eqnarray}
T&=&\frac{\left(25-c_M\right)}{24\pi}R\\
&=&\frac{\left(25-c_M\right)}{24\pi}e^{-\phi}\left(\overline{R}-\overline{\Delta}\phi
\right)
\end{eqnarray}

This trace can be derived from an effective action $S_{\rm eff}(\phi)$ via
\beq
T=\frac{2}{\sqrt{g}}\frac{\partial}{\partial\phi} S_{\rm eff}\left(\phi\right)
\eeq
We then find $S_{\rm eff}\left(\phi\right)=S_L\left(\phi\right)$.
The true action is then a sum of the classical and Liouville actions and
contains all the information on gravitational dressing, Hausdorff
dimension and baby universe structure.

The form of $S_L\left(\phi\right)$ has another consequence - it admits
{\it spike} solutions of the form $\phi_S\sim -\ln{r}$. These possess
an action which depends logarithmically on the IR cutoff $L$. They correspond
to quasi 1D geometries termed branched polymers.
A KT-like
argument then suggests that such configurations dominate $Z$ if
$c_M>1$ \cite{cates}. In this regime Liouville theory fails to give an adequate
description of the consequences of gravitational fluctuations.   

\section{Equivalence to DTs?}

Dynamical triangulations furnish another approach to quantum gravity in which
a finite simplicial mesh with invariant edge-length cutoff
is used to approximate the continuum
geometries. Summing over such lattices is hypothesized to
generate the correct measure on the space of geometries in some
suitable scaling limit. This conjecture is supported by the remarkable
fact that all correlation functions computed 
within Liouville theory agree with those coming from the DT ensemble 
for arbitrary genus surfaces \cite{amb}.

The question we would like to pose is whether it is possible to
demonstrate this connection between LT and DTs explicitly - initially
in 2D and later within the wider context of 4D models based on
quantizations of the conformal factor.

To answer this question within the context of numerical simulation
(the natural technique for exploring the DT models) we have devised a
procedure to reconstruct a lattice conformal mode for a generic
lattice drawn from a DT ensemble. We simply solve the
equation
\beq
R=e^{-\phi}\left(\overline{R}-\overline{\Delta}\phi\right)
\label{eq2d}
\eeq
For $\overline{g}$ we take a round sphere with fixed radius. Precise
lattice analogs exist for the Laplacian operator $\Delta$ on a DT geometry and
for the curvature scalar $R$. This is a nonlinear matrix equation which
we solve using an algorithm based on Newton iteration.

\subsection{Points to note}
On the basis of the Gaussian model LT predicts
\beq
\left<S_L\left(\phi\right)\right>=\frac{48\pi}{\left(25-c_M\right)}
\eeq
We have checked this for pure gravity $c_M=0$ and found reasonable
agreement. Fig 1. shows a plot of the mean Liouville action 
versus lattice size $V$. 

\begin{figure}[htb]
\vspace{9pt}
\epsfxsize=2.8 in
\epsfbox{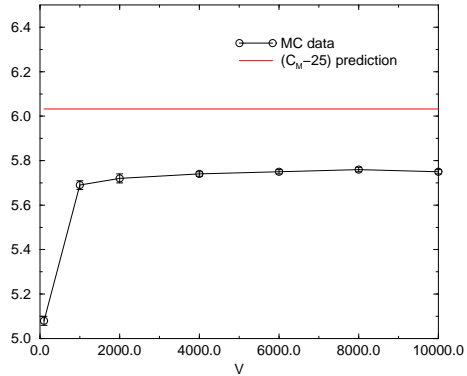}
\caption{Mean Liouville action versus lattice size}
\end{figure}

We have also examined the distribution of the lattice conformal
mode $P\left(\phi\right)$ as a function of $c_M$ (Figs. 2 and 3). We see that
$P\left(\phi\right)$ develops a broad tail for $c_M>1$
indicating the increasing dominance of {\it spikes} with
increasing central charge.

\begin{figure}[htb]
\vspace{9pt}
\epsfxsize=2.8 in
\epsfbox{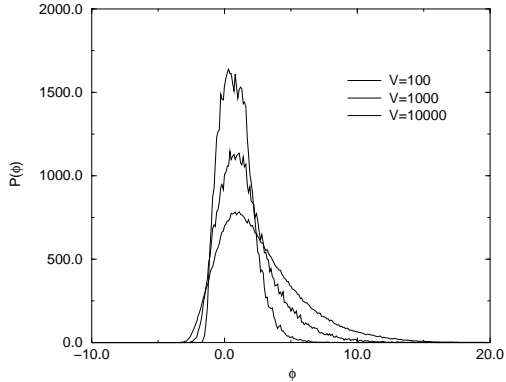}
\caption{Distribution of $\phi$ for $c_M=0$}
\end{figure}

\begin{figure}[htb]
\vspace{9pt}
\epsfxsize=2.8 in
\epsfbox{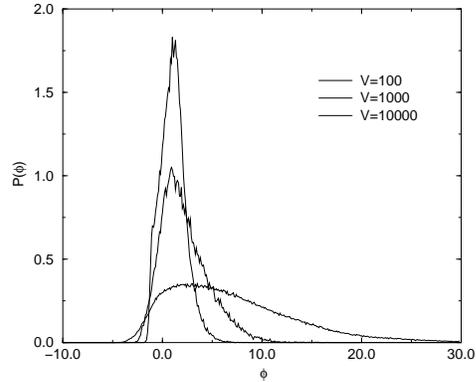}
\caption{Distribution of $\phi$ for $c_M=10$}
\end{figure}

\section{Analog 4D theory}

Consider again the conformal decomposition $g=\overline{g}e^{2\sigma}$. As
a first approximation we can consider fixing
$\overline{g}$ in order to study the quantum dynamics of the conformal
factor $\sigma$. This is expected to be a reasonable approach in the
I.R where for example the divergences encountered in the graviton
propagator stem from the spin zero sector. We can treat the
I.R dynamics of $\sigma$ exactly by analogy with the 2D case -- 
from the form of the trace of the energy momentum tensor
we can write down an effective action whose
variation yields the anomaly \cite{emil}. 
In conformal coordinates this yields an effective action
\beq
S_{\rm eff}\left(\sigma\right)=\frac{Q^2}{\left(4\pi\right)^2}\int \sqrt{\overline{g}}\left(\sigma\Delta_4\sigma+\frac{1}{2}
\overline{M}\sigma\right)
\label{eq4d}
\eeq
\beq
\overline{M}=\left(\overline{G}-\frac{2}{3}\overline{\Delta}\overline{R}\right)
\eeq

Here, $\Delta_4$ is the unique, fourth-order derivative operator which
transforms covariantly under conformal transformations and the parameter
$Q^2$ is the analog of the central charge in 2D -- depending on the
numbers of massless scalar, fermion and vector degrees of
freedom. 
\beq
Q^2=\frac{1}{180}\left(N_S+\frac{11}{2}N_{WF}+62N_V-28\right)
\eeq
Th quantity $\overline{G}$ is the Gauss-Bonnet density for the background
metric.
It is possible to absorb the contributions of the transverse gravitons into
an additive renormalization of $Q^2$ - $Q^2\to Q^2+Q^2_{\rm graviton}$.
Again, {\it spike} solutions
are possible and one expects two phases; a smooth Liouville-like
phase and a branched polymer phase caused by a condensation of
spikes. Notice though that in 4D the addition of more matter fields
{\it decreases} the influence of the spike-like configurations. 
Such branched polymer configurations have been seen in simulations
of 4D DTs and have led to the speculation that these represent
regularizations of this conformal factor gravity. If so, it should
be possible to enter the Liouville-like phase by the addition of
suitable massless fields -- preliminary numerical evidence indeed
hints at this \cite{biel,jap}. If this were so the simulations could be used to
both measure $Q^2_{\rm graviton}$ and to do nonperturbative studies of
relevance to cosmology \cite{emil2}.

To investigate this further we propose to follow our 2D calculations
and solve the analagous equation to eqn.~\ref{eq2d}
\beq
-\Delta\sigma+g^{\alpha\beta}\partial_\alpha\sigma\partial_\beta\sigma=
\frac{R}{6}-\frac{\overline{R}}{6}e^{-2\sigma}
\eeq

If this is done for an entire DT ensemble of 4-geometries we can ask
whether the resulting distribution for the lattice conformal mode
is again consistent with a simple gaussian action like eqn.\ref{eq4d}. 
This work is in progress.

\end{document}